\documentclass[aps,prd,eqsecnum,showpacs,amsmath]{revtex4}
\input epsf

\newcommand{\beq}{\begin{equation}}
\newcommand{\beqa}{\begin{eqnarray}}
\newcommand{\eeq}{\end{equation}}
\newcommand{\eeqa}{\end{eqnarray}}

\newcommand{\lsim}{\lesssim}
\newcommand{\gsim}{\gtrsim}

\newcommand{\vect}[1]{\mbox{\boldmath${#1}$}}
\newcommand{\lmk}{\left(}
\newcommand{\rmk}{\right)}

\newcommand{\ver}{{\vect r}}

\newcommand{\ven}{\vect n}
\newcommand{\vev}{\vect v}

\begin{document}

%\baselineskip 7mm
%\if0
%\draft
\title{Gravitational Wave Astrometry for Rapidly Rotating Neutron Stars
and Estimation of Their Distances} 
\author{Naoki Seto
}
\address{Theoretical Astrophysics, MC 130-33, California Institute of Technology, Pasadena,
CA 91125
}

%\fi
\begin{abstract}
We discuss an astrometric timing effect on  data analysis of
 continuous 
 gravitational  waves from  rapidly rotating isolated neutron
 stars. Special 
 attention is directed to the possibility of determining their distances
  by measuring the curvature of the wave fronts. We 
 predict that  
 if  
continuous  gravitational waves from an unknown
 neutron star with a stable rotation are detected  around 1kHz within
 $1/3$yr by initial 
 LIGO detectors and   the
 ellipticity parameter  $\epsilon$ is smaller than $10^{-6}$, the
 distance $r$ to the source can be estimated with 
 relative error 
 $\Delta r/r$ of $\sim 10\%$  by using the broad band configuration
 of  advanced LIGO detectors over 3 years.  
By combining  the observed  amplitude
 of the waves with the estimated distance,   information on the
 parameter $\epsilon$ can be obtained  purely through  gravitational 
 wave measurements. 

\end{abstract}
\maketitle

\section{Introduction}
Over the last   5 years, several large-scale interferometric
gravitational 
wave 
detectors, US LIGO \cite{Abramovici:1996dz}, Japanese TAMA
\cite{Ando:2001ej}, and British-German GEO \cite{Willke:2002bs}, were
constructed 
and have given us scientific data.  Italian-French VIRGO
\cite{Giazotto:1988gw} will be soon  
 in operation. In the next decade,  second generation
detectors 
({\it 
e.g.}  advanced LIGO \cite{adv},  LCGT \cite{Kuroda:2003nn})  are
expected to be  available.  More
ambitious third generation detectors ({\it
e.g.} EURO \cite{euro})  will  also be realized in the future. While the
current central 
objective of   scientific runs is the 
direct detection of  gravitational waves in 10Hz to a few kHz band, 
gravitational wave astronomy will be rapidly established 
with the advent of  new and  more sensitive detectors
\cite{Thorne_K:1987,Schutz:1999xj}. 
Once   gravitational waves from a population of sources are detected
 with relatively small signal-to-noise ratios (SNRs),   
 follow-on detectors would reveal  new astronomical information
at high SNRs. This kind of consideration is important for the
design 
and operational strategy in planing  next generation detectors.

 In this paper, we study an
 astrometric effect relevant to the data analysis of gravitational waves
from    rapidly
rotating isolated neutron stars and discuss the possibility of  estimating 
their distances.
%Astrometry is a basic technique  in
%astronomy  and provide us a way to estimate the distance  to  a
%object.
If gravitational waves from a rotating neutron star are detected, the
amplitude contains  information about 
the distance $r$ and  the ellipticity parameter
$\epsilon$ through the combination  $\epsilon/r$.  The parameter
$\epsilon$ characterizes the non-axisymmetry of the star.  However, it is 
  not easy to 
separate these two quantities individually from gravitational waves.
Here, we show  that by a  
long term observation of  gravitational waves from a 
source,  we can measure a 
small phase shift induced by an  astrometric effect.  
This means that  a rapidly rotating neutron stars are the ideal targets
to apply the long-term astrometric studies with  planned 
ground based detectors. Gravitational waves from  merging
neutron stars or 
 supernovae would last at most  order of seconds
\cite{Thorne_K:1987,Schutz:1999xj} and are not suitable for detecting
the shift.   
It is also interesting to understand how the quality of the astronomical
information depends on the observational duration. 
%For a very short
%phenomenon,  
%only the event rate is   relevant  to  the observational time.  
For a very short
phenomenon,  
 the only information improved by a long data span is the  event rate.

To
confirm  gravitational waves from a  rotating neutron star,
we have to integrate the signal for a  long time to get a sufficient
signal-to-noise ratio \cite{Brady:1997ji,Jaranowski:1998qm} (see
\cite{Abbott:2003yq} for a recent observational study with GEO and LIGO).
 The required computational
cost for the data analysis depends strongly on whether the neutron star is
known or unknown. If its location is unknown, we need a large number of
templates, and the data analysis is a 
very  challenging task \cite{Brady:1998nj, Cutler:2005pn}.
 The astrometric effect is larger for a closer
object, and   we implicitly assume a nearby
unknown neutron star  as our main  target.
Once  gravitational waves from an unknown neutron star are detected,  we
might identify an electromagnetic-wave counterpart using 
information 
  obtained from the  gravitational wave measurement,  and
estimate its distance with various traditional  methods (see {\it e.g.}
\cite{Gomez:2001te}).  
But a search for an  electromagnetic-wave  counterpart would be very difficult for  a cold
neutron star with 
a large peculiar velocity and   a very weak
magnetic field ({\it e.g.} less than $\sim 10^7$gauss critical  for
millisecond pulsars \cite{pk94}).

This paper is organized as follows:
In the next section we formulate the phase shift  induced by the
astrometric effect for a rapidly rotating neutron star, and explain a
way to estimate its distance though the  phase shift. Then, using the
Fisher matrix approach,  we obtain
an expression  for the magnitude of error associated with this estimation.
 This expression is
our central result.   In section III we apply it to 
 observational situations  with  specifications for
future projects, such as, 
advanced LIGO. We examine required conditions for the
ellipticity $\epsilon$ and the source distance $r$ in order to detect 
gravitational waves from rotating neutron stars and to determine
 their distances with the proposed method.  Then we   statistically study
whether there are nearby neutron stars to apply our
method. Section IV is devoted to a discussion including a brief summary.

\section{formulation}

We first study how well  we can extract  information on the distance
from  the phase modulation of  gravitational waves. Similar
arguments 
have been made  for  radio pulsar timing analyses
\cite{msec,msec2}, but 
this is 
the first 
attempt at applying  it  to  gravitational wave data.  Our
basic aim is   
to develop a model  relevant for evaluating  parameter
estimation errors.   We do not try to make a detailed model required
for  actual signal analysis {\it e.g.} including general
relativistic time delays in the solar system that are small and
introduce no additional   source
parameters \cite{Jaranowski:1998qm}. In addition, the
effects  of diurnal rotation are not  included here,  because we are
dealing with the case in which the signal
integration is of order a few years. 
Following Ref.\cite{Jaranowski:1998qm}, we use the solar system
barycenter (SSB) rest frame and investigate the 
effects induced by the motions of the detector as well as the source. We denote
the time dependent position of the detector $\ver_d$ ($r_d=|\ver_d|\sim
1$AU), and assume that the  
source is moving with a  constant velocity $\vev_{ns}$ as
$r\ven_0+\vev_{ns} (t+r/c)$ 
in the SSB frame. 
Here $r\ven_0$  is the position
of the source ($\ven_0$; source direction) from which the emitted wave
front reaches the SSB at 
$t=0$.   We define
$\ven_0=(\sin\theta\cos\phi,\sin\theta\sin\phi,\cos\theta)$ in the
spherical SSB coordinate  with
$\theta$ being the angle between the source direction $\ven_0$ and  the
normal 
vector of the  orbital plane (ecliptic) of the Earth. We write the
gravitational  
wave phase $\Psi_{SSB}$ observed at the SSB in the form
$\Psi_{SSB}(t)=\Phi_0+2\pi \sum_{k=0}^3 f_k t^{k+1}/(k+1)!$  and
perturbatively include the phase modulation due to the detector's and
source's motions 
with two small expansion parameters; $\alpha=|\ver_d/r|$ and
$\beta=|\vev_{ns} 
t|/r$. After some algebra, the phase   $\Psi_d$ observed by the detector
is 
given 
to $O(\alpha^m \beta^n)$ $(n+m\le 2)$  as 
\beqa
\Psi_d&=&\Phi_0+2\pi \sum_{k=0}^3 f_k \frac{t^{k+1}}{(k+1)!}
+\frac{2\pi r}{c} \lmk \ven_0\cdot \frac{\ver_d}{r} + \frac{\vev_{ns
\bot }t}{r} \cdot
\frac{\ver_d}{r}-\frac1{2r^2}(1-(\ven_0\cdot\ver_d)^2) 
  \rmk \sum_{k=0}^3 f_k \frac{t^{k}}{k!}\nonumber\\
& &+{\rm higher~order~terms},\label{phase}
\eeqa
where $\vev_{ns \bot }$ is the velocity component of $\vev_{ns}$
perpendicular to the direction $\ven_0$ \cite{msec}.
There are no  terms of order $O(\beta)$ and $O(\beta^2)$
\cite{Jaranowski:1998qm}.  
In the large parenthesis in the above equation, the $O(\alpha)$ term is
nothing but the plane wave effect that has been analyzed in the
literature and can be used to determine the direction of the
source. This term is periodic with a $1{\rm yr^{-1}}$ frequency. The
$O(\alpha \beta)$ term was discussed to some extent for  
 gravitational wave data  analysis in
Ref.\cite{Jaranowski:1998qm,Jaranowski:1998ge} and provides 
information on the proper motion of the source 
on the sky.   The
distance $r$ to the source is estimated through the 
$O(\alpha^2)$ term.  This term  probes the curvature of the
wave fronts as they deviate from  plane wave propagation. In the
$O(\alpha^2)$ term, the 
constant contribution $-1/(2r^2)$ is effectively absorbed in the
 phase constant $\Phi_0$, while the time dependent
oscillating term (with a $2{\rm yr^{-1}}$ frequency)  is proportional to 
$\sin^2 \theta$.  The order of magnitude of the time shift induced by
the plane wave $O(\alpha)$
term  in one year is written as $\sim r_d \cos \Theta/c\sim 500$sec with $\Theta$
being  the
angle 
between the two vectors $\ven_0$ and $\ver_d$. This angle
changes by $\Delta 
\Theta \sim r_d/r$ in one year due to  the parallax. Therefore the
$O(\alpha^2)$ term $\sim r_d^2/cr\sim  10^{-4}(r/30{\rm pc})^{-1}$sec 
can be regarded as a time shift induced 
by the 
parallax.  Even at $f\sim
1$kHz, this time shift produces a phase shift much smaller than the 1/4
cycle that is  critical for detection of gravitational waves, unless
the  
source is very close to the solar system. Therefore the distance $r$
cannot be 
estimated without a high SNR observation well beyond the required level for
gravitational 
wave detection.

Using a  model $h=h_c \sin[\Psi_d]$
($h_c$: constant) for the gravitational wave signal,   we calculated the
Fisher matrix for the following  10 fitting
parameters $\{f_0,f_1,f_2,f_3,\Phi_0,\vev_{ns,\bot},r,\theta,\phi\}$
and examined the relative estimation error $\Delta r/r$ with various sets
of   input parameters as well as the total observational time
$T_{obs}\gsim 
1$yr.
We found that the result $\Delta r/r$ is 
independent of the 
parameters  $\phi$, $\Phi_0$
and   inputs $\{f_1,f_2,f_3,f_4,\vev_{ns\bot}\}$, as long as these
inputs are close to reasonable values for real neutron stars.
  For a integration time $T_{obs}$ longer
than $\sim2$yr,  the resolution $\Delta r/r$ 
 is
approximated  well by the 
following expression 
\beq
\frac{\Delta r}{r}=0.11 \lmk \frac{r}{100{\rm pc}} \rmk \lmk
\frac{SNR}{500}  \rmk^{-1}
\lmk\frac{\sin \theta}{\sin\pi/3}  \rmk^{-2} \lmk \frac{f}{\rm 1kHz} \rmk^{-1} .\label{dr}
\eeq
We take $\theta=\pi/3$ for a reference value for the angle $\theta$, as it
bisects the area of a half sphere $0\le \theta \le \pi/2$. With the
specific 
choice of  parameters  $\theta=\pi/3$, $f=1$kHz,  $r=100$pc
and $SNR=500$, the prefactor 0.11 in eq.(2) becomes 9.14, 0.13, 0.109,
and 0.107 for 
$T_{obs}=1,2,3$ and 4 yr respectively. This indicates that  for $T_{obs}\gsim 2 $yr the
correlation between $r$ and other parameters decreases significantly and
becomes nearly stationary.
The coefficient becomes 0.107, when we remove $f_3$ or $\vev_{ns\bot}$
from the fitting parameters with
% the same $\theta$, $f$, $r$, SNR and
a fixed observation period  $T_{obs}=3$yr. In our model we have not
directly dealt with the effects of the acceleration of the
source. However, these  numerical experiments imply that, even if we
include  acceleration in a relatively simple manner, 
it would not change the resolution $\Delta r/r$ significantly.   When the
source is a
binary ({\it e.g.} with its orbital period $\sim 1$yr),  the
situation 
could change considerably. 

\section{Prospects with proposed detectors}

Next, we  discuss how well we can estimate the distance to a rotating
neutron star with the planned gravitational wave
detectors, such as,  advanced LIGO. 
We first describe the standard detection criteria of its gravitational
wave signal based on SNR for a matched filtering method \cite{Brady:1997ji,Brady:1998nj}.  Then we use a expression for SNR to  evaluate
the distance estimation error given by eq.(\ref{dr}).  For these
studies, we 
pay a close attention to  dependence of the ellipticity parameter $\epsilon$ and
the distance 
$r$ to  the source.

If the rotation axis of the neutron
star is identical to one  
of the principal axes of the inertial moment tensor $I_{ij}$, the
orientation averaged 
amplitude of  gravitational waves  becomes $h_c=8\pi^2
G\sqrt{2/15}If^2\epsilon/c^4r$, where  gravitational ellipticity is
defined as
$\epsilon\equiv 
(I_{22}-I_{11})/I\ll 1$, and  $I=I_{33}$ ($x_3$: the rotation axis) \cite{Thorne_K:1987}. 
With typical numerical values, we obtain 
\beq
h_c=7.7\times 10^{-26} \lmk\frac{\epsilon}{10^{-8}}
\rmk\lmk\frac{I}{\rm 10^{45} g~cm^2}  \rmk\lmk\frac{\rm 100pc}{r}
\rmk\lmk\frac{f}{\rm 1kHz}\rmk^2.\label{hc}
\eeq
Hereafter we fix the inertial moment at $I=10^{45}{\rm g~cm^2}$ and the
gravitational wave frequency at 1kHz. When the rotation axis is
different from a principal axis, the star precesses and emits
gravitational waves at
multiple frequencies \cite{Thorne_K:1987,Zimmermann:1979ip}. In this paper we use the above equation (\ref{hc})
as a reference to relate 
the characteristic amplitude $h_c$ with two parameters  $\epsilon$ and
$r$. 
For the sensitivity of  initial and  advanced LIGO at 1kHz, we
adopt the 
numerical values given in 
Figure 1 
in Ref.\cite{adv} (see also \cite{Buonanno:2001cj}).  We take $h_d=1.5\times 10^{-22}{\rm Hz^{-1/2}}$ for
a single 
4km detector of 
initial LIGO  and $h_d=4.2\times 10^{-24} {\rm Hz^{-1/2}}$ for 
advanced 
LIGO with  a
broad band configuration. After averaging with respect to  the direction
of the source, the effective sensitivity for a  monochromatic signal
becomes $h_{eff}=h_d\sqrt{2^{-1}5T^{-1}_{obs}}$ for an observational
period  
$T_{obs}$. The factor $2^{-1/2}$ comes from the number
of 4km detectors 
and another factor $5^{1/2}$ is due to the angular average with respect to
the source 
direction.    
We also study the case for the  proposed xylophone-type EURO detector
\cite{euro} with
$h_d=3.6\times 10^{-26}{\rm  
Hz}^{-1/2}$ at 1kHz.
 With these concrete specifications, the signal to noise ratio is calculated 
as  
\beq
SNR=\frac{h_c}{h_{eff}},\label{snr}
\eeq
 and the
resolution $\Delta r/r$ is now given as a function of $r$ and
$\epsilon$ from
eqs.(\ref{dr})(\ref{hc}) and (\ref{snr}). 

The characteristic SNR threshold required for detecting an unknown
rotating neutron star is discussed in \cite{Brady:1997ji}  and we take $SNR=20$ as a
reference value for  a $1/3$yr integration.  
In figure 1 we plot the $r-\epsilon$ relation for four observational
conditions;
$SNR=20$ with (i)  initial LIGO and (ii)  advanced LIGO  for
$1/3$yr integrations, and
 $\Delta r/r=0.1$ with (iii)  advanced LIGO and  (iv) 
xylophone-type  
EURO   for 3yr integrations.  
 We have $r\propto \epsilon$ for a given SNR,  but $r\propto
\epsilon^{1/2}$  for a given resolution $\Delta r/r$. If the parameter $\epsilon$ is fixed, we have
$SNR \propto r^{-1}$ and $\Delta r/r\propto r^2$. 
%We will take a close look at figure 1. Suppose the ellipticity parameter
%is $\epsilon=10^{-7}$.  From $1/3$yr data of  initial LIGO
%(advanced LIGO) we can detect
%gravitational waves  from
%neutron stars within $\sim 40$pc ($\sim 1300pc$) from the earth.
%By 3yr operation of advanced LIGO, we can also determine the distance
%to a neutron star with 
%error less than 10\% level, if it is within $\sim 150$pc.

To detect gravitational waves from neutron stars, a longer
integration time is statistically
advantageous.  This is because of the fact that
 the amplitude of the detectable signal decreases and  
the effective survey volume (or equivalently the event rate)
increases.
However, the  number of templates required for the matched filtering
method 
depends strongly on the 
integration time, and a manageable data length is limited
by computational 
resources. 
In reality, it might be difficult to deal with data  of  1/3 years,
especially 
for a   young neutron star with large coefficients $f_k$ ($k\ge 1$) (see eq.(\ref{phase})).
In this case,  we can detect neutron stars that are
closer than  the distances shown by the dotted lines in figure 1. 
However, with a  total  $\sim 10$ Teraflops
computational power  and
efficient searching algorithm, the differences between the dotted lines
and the maximum detectable   
distances for initial and advance LIGO would be less than a factor of 2 for most
neutron stars (see figure 5 in \cite{Brady:1998nj}). Once
gravitational wave from a neutron star is confirmed, it would be a
relatively easy task to analyze  longer data streams with a finer spacing
of 
template parameters. But  the available data span would not be much
larger than 3 years, considering the time scale of  major upgrades to 
next generation detectors.  This is the reason why we studied the case
with  a 3yr 
integration  for a follow-up
analysis of the 
distance measurement.  

 In figure 1, the solid line for advanced LIGO  and
the 
dotted 
line for initial LIGO intersect  around the interesting value $\epsilon =10^{-6}$,  close
to the upper 
limit from   theoretical  predictions
\cite{Brady:1997ji}. 
Almost all the triangle region in $(\epsilon,r)$-plane bounded by
$\epsilon \le 
10^{-6}$ (dashed line) and  the dotted line for initial LIGO is below the
sold line for advanced LIGO.
From  viewpoints of  future prospects of the gravitational wave
astronomy, this fact can be rephrased as follows: if an unknown neutron
star is detected around 1kHz 
with  
initial LIGO by $\sim3^{-1}$yr signal integration and  the ellipticity
has a  
reasonable 
magnitude $\epsilon \lsim 10^{-6}$,  advanced LIGO can determine its
distance  
with error of $\sim 10\%$ level within 3
years.  This is quite encouraging because blind
searches for  neutron stars are  very hard tasks requiring a huge
computational cost, as mentioned earlier \cite{Brady:1998nj}. Once it is
detected with  initial LIGO,  advanced 
LIGO would 
provide  information about its distance, one of the  fundamental
 astronomical parameters for a source.   The ellipticity parameter
$\epsilon$ may then also be found through knowledge of
the  amplitude $h_c$.  Both parameters are thus determined  purely
through  gravitational 
wave measurements. Note  that    advanced LIGO  has the  potential to
measure 
the distance to an unknown neutron star that cannot be  detected by initial
LIGO.  The effective sensitivity of the single EURO  xylophone-type
detector is 82 times better than  advanced LIGO detector at 1kHz, and
the 
distance $r$ for a given resolution
$ \Delta r/r$ is 9.1 times larger.

\begin{figure}
 \begin{center}
 \epsfxsize=9.cm
 \begin{minipage}{\epsfxsize} \epsffile{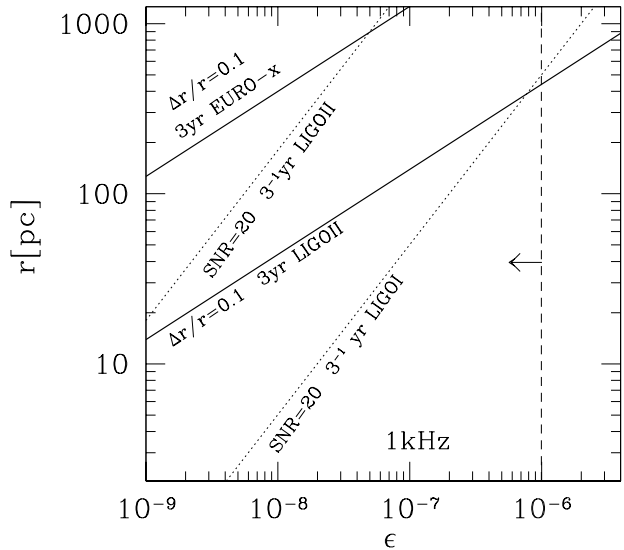} \end{minipage}
 \end{center}
\caption[]{ Relation between the distance $r$ and the gravitational
 ellipticity $\epsilon$ for given observational conditions of an
 unknown neutron star with gravitational wave frequency 
 $f=1$kHz.  The thin dashed lines represent the maximum distance for
 detection of gravitational waves with
 SNR=20 using  initial LIGO and  advanced LIGO for $1/3$yr integrations. The
 thick solid lines are the maximum distance for  resolution  $\Delta r/r=0.1$ 
 through the timing parallax method, shown for  advanced LIGO and 
 xylophone-type EURO for 3yr integrations. The dashed line $\epsilon =10^{-6}$ is  an
 upper limit of
 theoretical predictions for the ellipticity parameter. }  
\end{figure}

We have discussed the case in which the gravitational wave data would be
integrated without a break ($100\%$ duty cycle) for $3$ years or so.
But our result for $\Delta r/r$ would be approximately valid for a
coherent analysis of 
intermittent data streams that are sampled with appropriate intervals
($\ll 1$yr) and in a sufficient span ({\it e.g.} 3 years). If the narrow
band 
configuration is available at the target frequency of a source,  we
can reach the same resolution $\Delta r/r$ with a   shorter
total integration time. For example, the sensitivity $h_d$  could be
improved by  a
factor $\sim 3$  at 1kHz with the narrow band configuration of
 advanced LIGO, and
the total integration time could be a factor $\sim 10$ times smaller
than for its broad band configuration.

From Figure 1, we can estimate the distance $r$ for the required
resolution $\Delta r/r \sim 0.1$ as a function of the parameter
$\epsilon$. Here let us statistically study whether there is at least
one rapidly 
rotating pulsar within this distance. 
For this purpose we
use the 
argument originally made by Blandford assuming a population of neutron
stars 
whose spin evolutions are primarily determined by  energy loss through
gravitational radiation \cite{Thorne_K:1987,Brady:1997ji}. 
We  define $\tau_B^{-1}$ as the formation rate of the population in our
Galaxy with an initial gravitational wave frequency higher than
1kHz.  The
magnitude of the time $\tau_B$ is highly uncertain but should be much
larger than 
$\tau_B\sim 100$yr corresponding to the Galactic supernova rate.
%Our goal here is to estimate the formation rate  $\tau_B$ with which we
%can expect to have at least one pulsar within the distance for the
%resolution $\Delta r/r=0.1$. 
We first evaluate the minimum distance
$r_{min}$ to a 
pulsar of this population,  %as a function of  $\tau_B$ and $\epsilon$,
and then 
compare $r_{min}$ with the distance allowed for the
resolution $\Delta r/r=0.1$.  If these two distances coincide, 
we can expect to have at least one pulsar with  resolution 
$\Delta r/r$ better than 0.1. Our goal here is to calculate the required formation
rate $\tau_B$ for this condition as a function of the ellipticity
$\epsilon$.

In  Blandford's
model, the amplitude $h_c$ from a
neutron star with a fixed distance becomes larger for a larger
$\epsilon$. On 
the other hand, the evolution time scale $\tau_{GW}$ becomes smaller for a
larger $\epsilon$, and the smallest distance $r_{min}$ would be larger.
Here the evolution time scale $\tau_{GW}$ is defined  as
\beq
\tau_{GW}=\frac{f}{\dot f}=\frac{5c^5}{32\pi^4GI\epsilon^2 f^4}=1.8\times10^8 \lmk\frac{\epsilon}{10^{-8}}
\rmk^{-2}\lmk\frac{f}{\rm 1kHz}   \rmk^{-4}  {\rm yr}.
\eeq
The gravitational frequency becomes smaller by $\sim30\%$ within a single
$\tau_{GW}$.  The time scale $\tau_{GW}$ becomes the age of the universe
$\sim 
10^{10}$yr at $\epsilon\sim 10^{-9}(f/{\rm 1kHz})^{-2}$.
For the spatial distribution of the population in our Galaxy, we assume a
uniform density in a cylinder with radius $R_{ns}=10$kpc and 
height $H_{ns}=1$kpc. If the parameter $\epsilon$ is in the range
$10^{-9} (f/{\rm 1kHz})^{2}<\epsilon<10^{-6}(\tau_B/{\rm
200yr})^{-1/2}(f/{\rm 1kHz})^{-2}$, we have
\beq
r_{min}=\lmk3 V \tau_B \tau_{GW}^{-1}/4  \rmk^{1/3}=44
\lmk\frac{\tau_B}{\rm  200 yr}  \rmk^{1/3}
\lmk\frac{\epsilon}{10^{-8}}  \rmk^{2/3}
\lmk\frac{f}{\rm 1KHz}  \rmk^{4/3}
{\rm pc} 
\eeq
for the smallest distance. The distance $r_{min}$ becomes larger than the
height $H_{ns}$ for $\epsilon>10^{-6}(\tau_B/{\rm
200yr})^{-1/2}(f/{\rm 1kHz})^{-2}$. In this case we have $r_{min}\sim
R_{ns} (\tau_B \tau_{GW}^{-1})\propto \epsilon^{1/2}$. With these
expressions and results presented in figure 1,  we can now estimate the required formation rate $\tau_B$
for observing at least one pulsar with resolution $\Delta r/r=0.1$. For
3yr advanced LIGO observation at 1kHz, we have 
$\tau_B<84  
(\epsilon/10^{-8})^{-1/2}$yr in the range $10^{-9}<\epsilon<10^{-5}$.
Therefore, a very small
$\tau_B$ (or a large formation rate 
$\tau_B^{-1}$) is required from this rough  statistical argument.
%Note that the time $\tau_B$ becomes unrealistically small for a large
%$\epsilon$. 
 In the
case of 
3yr EURO-xylophone-type data, we obtain $\tau_B<6.3\times
10^4(\epsilon/10^{-8})^{-1/2}$yr for $10^{-9}<\epsilon<
10^{-7}$.

So far,  we  have  studied the case with $\Delta r/r\ll 1$.
Here, we discuss  the opposite case in which the distance to a
source is large and we cannot detect the signature of the parallax
timing shift. 
In the case of  usual angular  parallax measurement, the amplitude of
the apparent annual motion of a source on the sky is inversely 
proportional to its distance $r$. Therefore, it is straightforward to
fit
  distance with a form $r^{-1}$ in the parameter estimation.  We can
understand this 
from the original meaning of the distance unit ``parsec'' that represents
$1/{\rm 
arcsec}$.  The situation is similar to the fitting for the parallax timing.
 Note also that the error $\Delta (r^{-1})\sim (\Delta 
r)/r^2$ for $r^{-1}$ does 
not depend on the distance $r$ itself, as shown in eq.(2). In the
 low resolution limit with a fitting result
$|(r^{-1})_{fit}|\sim \Delta (r^{-1})$,  we can set a constraint 
for the value $r^{-1}$ in the form
$r^{-1}<\max\{(r^{-1})_{fit},0\}+N\Delta(r^{-1}) $  with $N-\sigma$
statistical 
significance. 
At  the end we get a lower limit of the distance $r$ as  
\beq
r>(N+x)^{-1} \{\Delta
(r^{-1})\}^{-1}=220\lmk \frac{SNR}{500} \rmk \lmk
\frac{\sin\theta}{\sin(\pi/3)}\rmk^2  \lmk \frac{
f}{1{\rm 
kHz}}\rmk \lmk \frac{N+x}4\rmk ^{-1}{\rm pc}
\eeq
 for
$T_{obs}\gsim2$yr 
 with $x\equiv\max\{(r^{-1})_{fit}/\Delta
(r^{-1}),0   \}$.  We can also set a lower limit to  the parameter
$\epsilon$ by using the gravitational wave amplitude
$h_c$.

\section{Discussions}

In our analysis, the phase $\Phi_{SSB}(t)$ at the SSB is assumed to be
simple  and be described by a low-order Taylor expansion. 
Even though many pulsars are known to be extremely accurate clocks, the
phase $\Phi_{SSB}$ could have a deviation from this simple
expression. The 
origin of these timing deviations can be due to  the intrinsic evolution
of the pulsars or other effects ({\it e.g.} the low frequency
gravitational wave background).
%\cite{Seto:2005tq} 
If the deviation is  large,  especially around the
frequency $\sim2{\rm yr^{-1}}$, it would hamper the estimation of the
distance through  parallax timing even with a high SNR measurement. 
Here, we briefly describe the actual radio observation of certain
pulsars following \cite{msec3}.
 PSR B1855+09 is a binary  millisecond pulsar with  spin period
5.36 msec  
and  orbital period 12.3 day, and PSR B1937+21 is a single
millisecond pulsar with  spin period 1.56 msec. They are considered to be
recycled pulsars. Timing analyses for both pulsars have been performed
with data 
taken at  Arecibo Observatory in a  $\sim8$ year span.  A 
systematic timing deviation of order a few $\mu$sec is indicated for
PSR B1937+21, but not for PSR B1855+09. 
The timing residual of  the  former is expected to be dominated by the
intrinsic evolution of the pulsar and not by other effects, such as, the
low 
frequency gravitational wave background.
As for  the estimated distances
through the timing parallax, PSR B1855+09 has a fitted value
$\pi\equiv(1{\rm 
kpc}/r) {\rm mas}=1.1\pm0.3$mas corresponding to
$r=0.9^{+0.4}_{-0.2}$kpc (1$\sigma$), and  PSR B1937+21 has a bound
$\pi<0.28$mas (2$\sigma$) corresponding to $r>3.6$kpc.
Their distances are  also estimated from dispersion measures together
with a 
model of the Galactic distribution of  interstellar free
electrons. They  are 0.7kpc for PSR B1855+09 and 3.6kpc for PSR B1937+21
with uncertainties of $\sim25\%$, and are consistent with the
estimation through the timing parallax. Therefore, the effects of the
timing 
deviation would  be moderate in these cases.

Now we summarize this paper.  We  have discussed an astrometric effect
for the analysis of gravitational waves that would allow us to determine
the distance to a 
rapidly rotating isolated neutron star around $f\sim1$kHz. The resolution
$\Delta  
r/r$ improves 
significantly,  if the integration period of the signal is longer than
$\sim 
2$ years. By   
using 
 advanced LIGO detectors for 3 years at its designed sensitivity in
the 
broad 
band configuration, we can estimate its distance
with an error of $\Delta 
r/r\lsim 0.1$  up  to $r\sim
150(\epsilon/10^{-7})^{1/2}$pc, if its spin evolution is relatively
simple.

{\it Acknowledgments:} The author is grateful to Teviet Creighton 
 for valuable discussions.
He also thanks Sterl Phinney,
Kip Thorne and Yi Pan for their help and useful conversations, and
  Asantha Cooray  and  Steve Furlanetto for carefully reading the
manuscript. This work was 
supported by  NASA grant  NNG04GK98G and  the  Japan Society for the Promotion of Science.

\end{document}